\documentclass[doublecol, figures]{epl2} 
\usepackage{graphicx}
\usepackage{amsfonts}
\usepackage{amsmath}
\usepackage{epsfig}

\title{Disorder, Metal-Insulator crossover and Phase diagram in
high-$T_{c}$ cuprates}
\shorttitle{Disorder, Metal-Insulator crossover and Phase diagram in
high-$T_{c}$ cuprates} 

\author{F.\ Rullier-Albenque\thanks{E-mail: \email{florence.albenque-rullier@cea.fr}}\inst{1} \and H.\ Alloul\inst{2} \and F.\ Balakirev\inst{3}\and C.\ Proust\inst{4}}
\shortauthor{F.\ Rullier-Albenque \etal}

\institute{                    
  \inst{1} Service de Physique de l'Etat Condens\'{e} (CNRS URA 2464), CEA Saclay, 91191 Gif sur Yvette cedex, France\\
  \inst{2} Laboratoire de Physique des Solides, UMR CNRS 8502, Universit\'{e}
Paris-Sud, 91405 Orsay, France\\
 \inst{3} National High Magnetic Field Laboratory, Los Alamos National Laboratory, Los Alamos, New Mexico 87545, USA\\
\inst{4} Laboratoire National des Champs Magn\'{e}tiques Puls\'{e}s, UMR CNRS-UPS-INSA 5147, Toulouse, France\\}


\abstract{We have studied the influence of disorder induced by electron irradiation on
the normal state resistivities $\rho(T)$ of optimally and underdoped YBa$_{2}%
$Cu$_{3}$O$_{x}$ single crystals, using pulsed magnetic fields up to
60T to completely restore the normal state. We evidence that point defect disorder induces
low $T$ upturns of $\rho(T)$ which saturate in some cases at low $T$ in large
applied fields as would be expected for a Kondo-like magnetic response. Moreover the magnitude of the upturns is related to the residual resistivity, that is to
the concentration of defects and/or their nanoscale morphology. These upturns are found quantitatively identical to those reported in lower ${T_{c}}$ cuprates, which establishes the importance of disorder in these supposedly pure compounds. We therefore propose a realistic phase diagram of the cuprates, including disorder, in which the superconducting state might reach the antiferromagnetic phase in the clean limit.}

\begin{document}

\maketitle

\section{Introduction}
One of the most exotic aspects of superconductivity in correlated
electron systems remains its interplay with magnetism. In the so-called
"generic" phase diagram of the cuprates,
a dome shaped superconducting region is usually shown separated from an antiferromagnetic
phase which is insulating in the undoped compounds. A metal-insulator
crossover (MIC) has been initially assigned to the doping at which low
${T}$ upturns of the resistivity appear. 
Those usually occur as logarithmic divergences of the resistivity, i.e. $\rho(T)\sim-\log T$, that cannot be explained by usual electron localization theories.
This has been mainly evidenced in
low $T_{c}$ compounds ($T_{c}\leq40$K) La$_{2-x}$Sr$_{x}$CuO$_{4}$
(LSCO) \cite{Ando1, Boebinger}, Bi$_{2}$Sr$_{2-x}$La$_{x}$CuO$_{6+\delta}$
(La-Bi2201) \cite{Ono} and Bi$_{2}$Sr$_{2}$CuO$_{6}$ \cite{Vedeneev} for which the highest available magnetic field are sufficient to suppress superconductivity.  In the underdoped regime of these compounds  the  -$\log T$ behavior has been often
considered as an intrinsic feature induced by the applied field and interpreted as a signature of a competing order that coexists with superconductivity \cite{Sun1, Hawthorn}. As static antiferromagnetic (AF) order has been found to
be also induced by high magnetic field in LSCO \cite{Lake}, it has even
been suggested that the ground state of these compounds for
$H>H_{c2}$ would be an AF insulator.

On another hand, a zero field superconducting-insulator transition can also be induced in the cuprates by the introduction of Zn at the Cu site of the
CuO$_{2}$ planes, with a similar -$\log T$ dependence of the resistivity
\cite{Fukuzumi, Segawa, Hanaki}. This has been taken as an indication that the
 "insulating" behaviors induced by the
magnetic field or by disorder have a common origin \cite{Marchetti, Komiya}.
However as it is now often suggested that defects are at the origin of low
$T_{c}$ values \cite{Bobroff97,Eisaki}, it remains very difficult to
distinguish the influence of the applied field from that of disorder.

In this letter we address this issue by studying a clean compound in which defects were intentionally introduced in a controlled manner. As a starting material we have chosen YBCO, which has been shown by NMR to be a very clean compound \cite{Bobroff1}. We have further established recently that underdoped YBCO$_{6.6}$ displays a metallic behaviour 
under high magnetic fields down to 1.5K \cite{RA4}. We have previously shown that low $T$ electron
irradiation is a reliable means to introduce point defects, presumably oxygen
and copper vacancies, in the CuO$_{2}$ planes of the cuprates \cite{Legris,
RA-EL}. In YBCO$_{7}$ superconductivity is depressed in a quantitatively
similar manner to Zn substitution, without changing the hole doping. We show here that the introduction of defects by electron irradiation
in optimally (OP) or underdoped (UD) YBCO always induces a MIC which is better revealed by fully restoring the normal state with high magnetic fields. 
For low defect content, we find that the upturn contributions to
the resistivity display the main features expected for the Kondo effect in
dilute alloys. We reinforce then previous suggestions based on magnetic and
transport properties \cite{Bobroff2, RA1} and demonstrate that disorder is
responsible for the apparent "insulating" behavior. By
introducing nanoscale inhomogeneities, we observe that $\Delta\rho_{2D}$ still 
displays a $-\log T$ dependence but
with a much steeper slope than for point defects. The comparison with the results reported in "pure" low $T_{c}$ compounds leads us to suggest that the magnitude of the upturns is governed by the length scale of the disorder.

\section{Samples and techniques}
The single crystals were grown using the standard flux
method. Contacts with low resistivity were achieved by evaporating gold pads
on the crystals on which gold wires were attached with silver
epoxy. Subsequent annealings were performed in order to obtain crystals with
oxygen content around 7 and 6.6. The characteristics of the samples used in this study are described in table \ref{tab.1}.
 
The irradiations were carried out with MeV (usually 2.5MeV) electrons in the
low $T$ facility of the Van de Graaff accelerator at the Laboratoire
des Solides Irradi\'{e}s (Ecole Polytechnique, Palaiseau, France). The samples were immersed in liquid hydrogen and the electron flux
was limited to 10$^{14}$ e/cm$^{2}/s$ to avoid heating of the samples during
irradiation. The thicknesses of the samples were very small compared to the
penetration depth of the electrons, ensuring a homogeneous damage throughout.
After irradiation, the samples were warmed up to room $T$ then
cycled back to low $T$ and\ $\rho(T)$ was measured simultaneously
using a standard four-probe dc technique.

As long as $\rho(T)$ after irradiation and annealing at room
$T$ is identical to that observed in situ for lower irradiation
fluences, we can assert that the resistivity is only sensitive to the
concentration $n_{d}$ of randomly distributed in-plane defects. This has been the case for all samples of table \ref{tab.1} except O$_{6.6}$-C and D that will be discussed later.

Pulsed fields up to 60 T were used in Los Alamos (LA) or in Toulouse (T) with the field applied along the $c$ axis in order to better suppress SC. The durations of the pulses have imposed
the ac current frequencies of the measurements which vary from 10kHz for long
pulses ($\>$150ms), to 250kHz for short pulses (20ms). In the latter case, we have taken
particular care to control the absence of eddy current heating by comparing
the up and down field sweep data or by using pulses with different peak
fields.

\begin{table}[h]
\caption{Characteristics of the samples. The
residual resistivities $\rho_{0}^{2D}=2\rho_{0}/c$ ( where $c$
is the c-axis lattice spacing) have been determined by fitting the high-$%
T$ parts of the $\rho (T)$ curves as explained in the data
analysis. The defect content $n_{d}$ has been estimated from the increase of 
$\rho _{0}$ with electron fluence \cite{RA2}. Samples labeled O$_{6.6}$-A1 and O$_{6.6}$-B1 correspond to the same single crystal irradiated at two different electron fluences.}
\label{tab.1}
\begin{center}
\begin{tabular}{ccccccc}
\textbf{Sample} & \textbf{T}$_{c}$ & \textbf{$\rho_{0}^{2D}$} & \textbf{n$_{d}$} &  
\textbf{magnet}\\
O$_{7}$-A & 30K & $2.3k\Omega/\square$ & 3.9\% & T-60T-126ms\\
O$_{7}$-B & 1.9K & $4.8k\Omega/\square$ & 8.2\% & LA-45T-500ms\\
O$_{6.6}$-p & 57K & $0.6k\Omega/\square$ & - & T-60T-126ms\\
O$_{6.6}$-A1 & 25K & $3.7k\Omega/\square$ & 1.6\% & LA-60T-20ms\\
O$_{6.6}$-A2 & 25K & $3.7k\Omega/\square$ & 1.6\% & T-60T-126ms\\
O$_{6.6}$-B1 & 6.8K & $6.1k\Omega/\square$ & 2.8\% & T-60T-126ms\\
O$_{6.6}$-C & 3.5K & $8k\Omega/\square$ & $\sim\%4$ & LA-60T-20ms\\
O$_{6.6}$-D & - & $12.4k\Omega/\square$ & $\sim6\%$ & zero field\\
\end{tabular}
\end{center}
\end{table} 

\section{High field measurements and data analysis}
\begin{figure}[h]
\centering
\includegraphics[width=8cm]{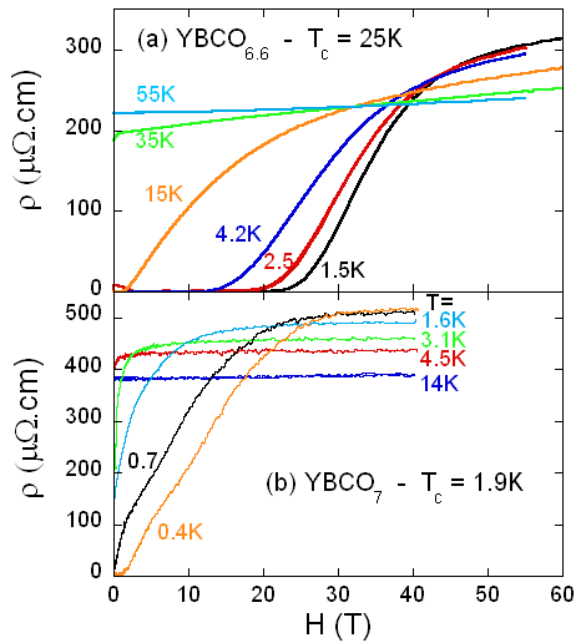}
\caption{Resistivity $\rho$ versus magnetic field at various temperatures for (a) YBCO$_{6.6}-$A2 and (b) YBCO$_{7}$-B.}
\label{fig.1}
\end{figure}

Fig.\ref{fig.1} shows the $\rho(H)$ curves for
two irradiated samples. One can see that a 60T magnetic field is hardly
sufficient to suppress SC below 15K in the UD sample with $T_{c}\sim 25$K
(fig.\ref{fig.1}-a) and that a 35T field is still needed for the OP sample
with $T_{c}\sim 2$K (fig.\ref{fig.1}-b).

The magnetoresistance in the highly disordered OP sample is nearly zero, so
the resistivity value at 40T can be taken as the normal state value for $%
T<T_{c}$. This is also valid for the most irradiated O$_{6.6}$-C sample.
However in some samples such as that of fig.\ref{fig.1}-a or O$_{6.6}$-B1, for which
the $\rho (H)$ curves were reported in \cite{RA4}, the high field magnetoresistance is not negligible. We have thus extrapolated the
high field data down to zero field using the fact that the
magnetoresistance increases as $H^{2}$ in the normal state \cite{RA4}. In
any case the $\rho (T)$ dependences obtained with extrapolated values or
directly from the 60T data do not differ significantly.

\begin{figure}
\centering
\includegraphics[width=8cm]{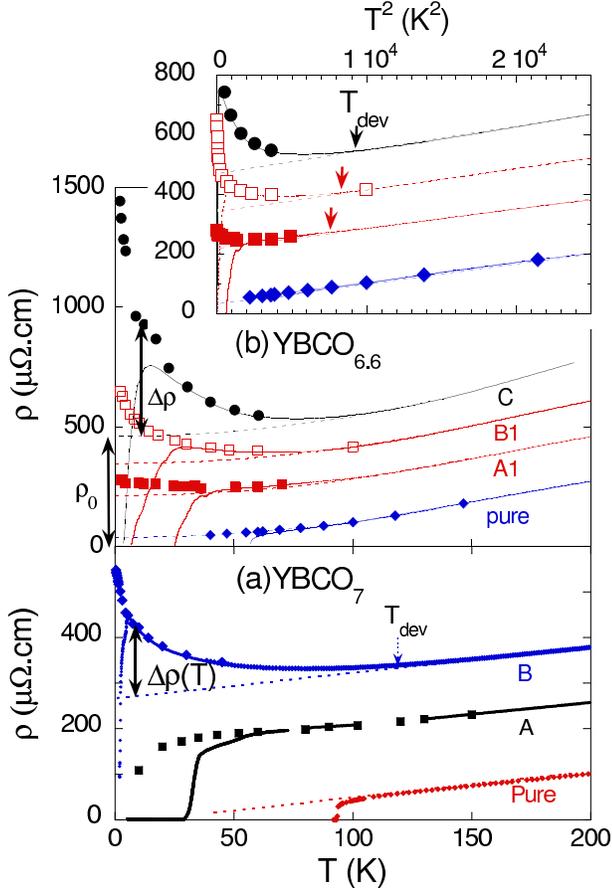}
\caption{$\rho(T)$ data taken in zero field (solid lines) and obtained from high field data by
extrapolation (as explained in the text):
(a) for O$_{7}$ samples. Dashed lines are linear extrapolations down to $T=0$ of the high-$T$ part of the resistivity.
(b) for O$_{6.6}$ samples with increasing defect content.
The $T^{2}$ plot of the inset allows us to emphasize the $T^{2}$ variation at high $T$. The arrows indicate the location of $T_{dev}$ above which $\Delta \rho (T)$ becomes negligible within experimental accuracy.}
\label{fig.2}
\end{figure}

In fig.\ref{fig.2} we display the $\rho(T)$ curves for the pure YBCO$_{6.6}$ and YBCO$_{7}$ in
zero field together with that obtained in all the irradiated samples
studied. For both OP and UD YBCO, the $\rho(T)$ curves are perfectly
parallel for $T\geq100$K, showing that the irradiation defects add a
$T$-independent contribution to the high-$T$ resistivity. This is an indirect
confirmation that the hole doping is not modified by the introduction of defects.
The low $T$ upturns of $\rho(T)$, already visible
in zero field in the most irradiated samples, are further disclosed down to
lower $T$ when SC is additionally suppressed by a field. The continuity of
the data gives evidence that \textit{the upturns are related to the defects and not
induced by the field}, even when an apparent 
"metallic" behavior is observed in zero field.
To perform a quantitative analysis we can write
\begin{equation}
 \label{Eq.1}
\rho(T)=\rho_{0}+\Delta\rho(T)+\rho_{i}(T)
\end{equation}
where $\rho_{0}$ and $\Delta\rho(T)$ are respectively the residual resistivity
and the upturn defect contribution and $\rho_{i}(T)$ is the $T$ dependent
resistivity of the host, which is seen to be nearly unaffected by disorder (so $\rho_{i}(0)=0$). In order to extract the defect contribution $\Delta \rho (T)$, it is
necessary to extrapolate the high-$T$ part of the resistivity down
to low $T$. In the case of O$_{7}$-B a linear dependence of $\rho
_{i}(T)$ was used to extract $\rho _{0}$ and $\Delta \rho (T)$. For YBCO$%
_{6.6}$ one can see in the inset of fig.\ref{fig.2} that the $T^{2}$ dependence of $\rho _{i}(T)$
which is observed in the pure crystal in the range $\sim 60-170K$ \cite{RA4}
is still observed above $\sim $100K in the irradiated samples. The $\rho%
_{0} $ values were determined by extrapolating this $T^{2}$ dependence down
to 0K, as indicated in fig.\ref{fig.2}. The corresponding data for $\Delta \rho (T)$ as obtained from eq.\ref{Eq.1} are found to become significant below temperatures $T_{dev}$ 
that slightly increase with increasing defect content ( arrows in fig.\ref{fig.2}).

\begin{figure*}[t]
\centering
\includegraphics[width=17cm]{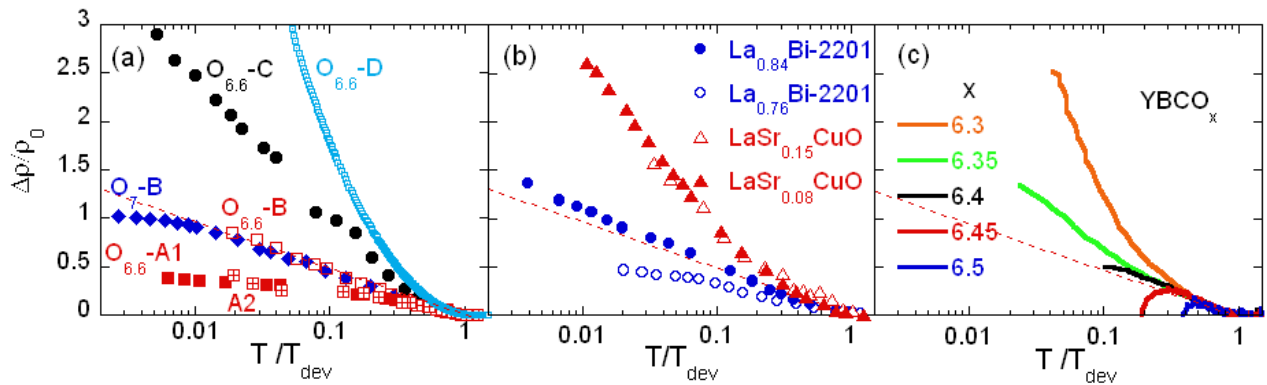}
\caption{Defect contribution to $\rho(T)$ normalized by the residual resistivity $\rho_{0}$ versus $T/T_{dev}$ $^{(*)}$
(a) for the irradiated YBCO samples  that display an upturn (symbols as in
fig.2). A clear MIC is only  visible on sample O$_{6.6}$-D in which superconductivity has been totally suppressed by irradiation.
(b) for LS$_{x}$CO and La$_{x}$Bi-2201. These data were taken from ref. \cite{Boebinger} with $T_{dev}=65$ and $70$K  respectively for $x=0.15$ and $0.08$ and from ref. \cite{Ono} with $T_{dev}=80$ and $25$K for $x=0.84$ and $0.76$.
(c) for pure UD YBCO$_{x}$ with $x$ ranging from 6.5 to 6.35 \cite{Ando3} and $x = 6.3$ \cite{Ando4}. Here $T_{dev}=115, 140, 149, 155$ and 160K respectively from 6.5 to 6.3. The dotted line, which is the high-$T$ fit of the -$logT$ behavior in panel (a), is repeated in panels (b) and (c) for comparison.
$^{(*)}$ Here the normalization by $T_{dev}$ helps to compare quantitatively the data for different samples}
\label{fig.3}
\end{figure*}

The values of $\Delta\rho(T)$ normalized to $\rho_{0}$ are plotted in fig.\ref{fig.3}-a
versus $\log T$. We have analyzed the results obtained in low $T_{c}$
cuprates LSCO \cite{Boebinger} and La-Bi2201\cite{Ono} as well as in the lightly doped YBCO$_{x}$ with $x$ ranging from 6.5 to 6.3 \cite{Ando3, Ando4} with the same method. We used a linear $T$ dependence for $\rho_{i}(T)$ in the case of LSCO at optimal doping and a $T^{2}$ dependence in the case of all the underdoped compounds. We obtain the results of fig.\ref{fig.3}-b and c which display a quantitative analogy with our data of fig.\ref{fig.3}-a. This allows us to conclude  that the resistivity upturns are related to $\rho_{0}$, that is to \textit{the presence of
disorder even in the "pure" cuprates}. As will be discussed below the differences between the cuprate families can be assigned   to distinct nanoscale disorder.

\section{"Kondo" analysis for low defect content}
Let us now discuss quantitatively the $\Delta\rho/\rho_{0}$ curves for irradiated samples labeled A and B in fig.\ref{fig.3}-a. They display
a -$\log T$ dependence at high enough $T$, but with a
downward deviation below $\sim30$K in samples O$_{6.6}$-A and O$_{7}$-B \cite{rem}. This
saturation cannot be related to superconductivity as we have seen in fig.\ref{fig.1} that the
applied field fully restores the normal state in these cases. Thus we can conclude that the resistivity upturns and saturations observed here are not signaling a \textit{metal-insulator transition} but are associated with \textit{an inelastic scattering on the point defects}.

\begin{figure}[h]
\centering
\includegraphics[width=8cm]{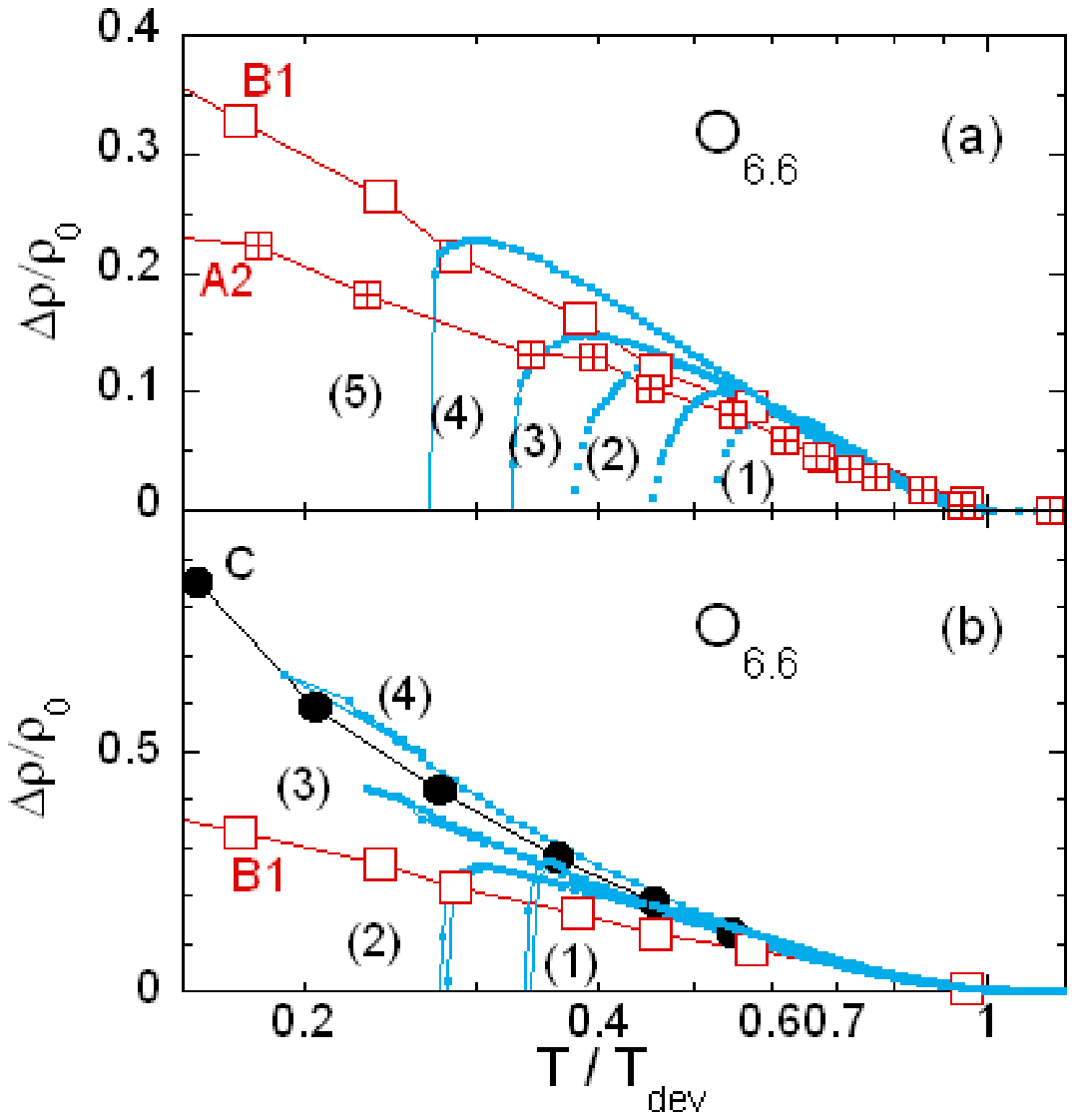}
\caption{Evolution of the defect contribution to the resistivity with defect
content and nanostructure in O$_{6.6}$ samples. (a) For increasing point defect content: comparison of the $\Delta\rho
/\rho_{0}$ data versus $T/T_{dev}$ obtained in high field in O$_{6.6}$-B$_{1}$, A$_{2}$
with those measured in-situ in zero field in another
O$_{6.6}$ sample (small blue squares). Those are labelled by increasing numbers
corresponding respectively to $\rho_{0}^{2D}$ values of 1.5, 2.2, 3.0, 3.8, 4.8
$k\Omega/\square.$
(b) For clustered disorder: comparison of the $\Delta\rho/\rho_{0}$ data
versus $T/T_{dev}$ obtained in high field in O$_{6.6}$-B$_{1}$ and C ,
with those measured in-situ in zero field in the same O$_{6.6}$ sample (small
blue squares) after annealing and further irradiation. Those are labelled by
increasing numbers corresponding respectively to $\rho_{0}^{2D}$ values of 3.3,
4.2, 6, 8.5 $k\Omega/\square$.}
\label{fig.4}
\end{figure}

It is therefore tempting to relate these effects with the Kondo like behavior
observed by NMR for the local moment susceptibility induced by spin-less
impurities like Zn or Li \cite{Bobroff2}. We had indeed shown \cite{RA1} that
for less irradiated O$_{6.6}$ samples the $\Delta\rho(T)$ taken in the
irradiation set-up scaled linearly with the defect content as long as
$\rho_{0}^{2D}<5k\Omega/\square$. In such zero field experiments it was
impossible to extract the actual $T$ dependence of $\Delta\rho$ because of the proximity of
$T_{c}$. When re-analyzing these data with the procedure used here to subtract
$\rho_{i}(T)$, we get the very good scaling of $\Delta\rho$ with $\rho_{0}$
displayed in fig.\ref{fig.4}-a for selected defect contents. Consistently, the scaling is good as well for $T/T_{dev}>0.4$ in two distinct samples in which the
$T$ variations are revealed by the applied field. 

All the features above: i) proportionality to defect content, ii) -$\log T$
dependence and iii) saturation at low $T$ in some cases are the well known
benchmarks expected for Kondo inelastic spin-flip scattering effects induced
by magnetic impurities in classical metals. In UD YBCO$_{6.6}$ for which the "Kondo" temperature is very low  \cite{Bobroff2}, the spin-flip
scattering should be suppressed by magnetic fields. However precise studies of
the magnetoresistance are not possible as it is difficult to separate  the suppression of superconductivity from the normal state field variation of $\rho$. The downward deviation from the -$\log T$ dependence observed below 30K in the O$_{6.6}$-A samples might be speculatively associated with a saturation of the magnetization, as it occurs for $\mu_{B}H\sim k_{B}T$ (effective moment of $\sim1\mu_{B}$ \cite{RMP}).
Let us note here that in\ underdoped BiLa2201 in which 2.2\%Zn totally
suppressed SC, a negative magnetoresistance has been reported, and interpreted
by a Kondo scattering due to the Zn induced local moments \cite{Hanaki}.
In the case of Li substituted optimally doped YBCO$_{7}$ samples, NMR studies have shown that
the susceptibility of the moments induced by Li impurities, which behave as for Zn, displays a $(T+\Theta)^{-1}$ variation, $\Theta\sim100K$ being possibly
associated with a Kondo temperature \cite{Bobroff2}. In that case one would
expect a saturation of the resistivity at $T\leq\Theta$. But attempts to
control this possibility could only be performed for a large concentration of
defects in order to suppress SC. Indeed the -$\log T$ dependence observed in
O$_{7}$-B displays a saturation around 2K and would be compatible with
$\Theta\approx10$K. Such a reduction of $\Theta\;$could be related to
strong interactions between defects (here $n_{d}\sim8\%$).

Compared to the Kondo effect found in classical metals, the situation
investigated here is somewhat unusual as the local moment formation and its
screening result from the same bare interactions \cite{RMP}. Recent calculations which
consider AF fluctuations in a Fermi liquid approach have shown that impurities
might induce "Kondo-like" upturns of
resistivity when the system is close to the AF quantum critical point
\cite{Kontani}. Although this approach might be questionable in underdoped
cuprates, it provides a first attempt to explain resistivity upturns induced
by spin-less impurities in these strongly correlated systems.

\section{Influence of the disorder morphology }
It is clear from the three panels of fig.\ref{fig.3} that the high $T$ logarithmic slope of $\Delta\rho(T)/\rho_{0}$ has the same magnitude in La-Bi2201 and in YBCO$_{x}$ as in our YBCO crystals with a moderate defect content. This leads us to suggest that point defects are responsible for the low $T_{c}$ of these compounds. As previously noticed \cite{Ono} the "insulating" behavior in LSCO samples is much stronger than in La-Bi2201. This could be associated with the formation of inhomogeneous states such as stripes and/or peculiar disorder dominated by tilts of the CuO octahedra \cite{Tranquada, Attfield}.

In order to clarify the role of the defect microstructure, we  have induced a clustering of the defects in sample O$_{6.6}$-C. This sample was irradiated by
1.8MeV electrons at a fluence which allowed to fully suppress its $T_{c}$.
Annealing at high $T$ (400K) that favors defect migration on large distances, results in clustering of defects. Meanwhile superconductivity was recovered with $T_{c}\sim3.5$K. One can see in fig.\ref{fig.4}-b that the $\Delta\rho(T)/\rho_{0}$ curve
departs from the dependence observed in the other samples towards a new -$\log T$ with a slope about three times larger and no sign of saturation down to 0.5K.

To confirm the generality of this observation, we have considered an UD sample which had been aged at room T for two years after irradiation, so that clustering of defects did occur. In such a sample further irradiation does not simply introduce additionnal random point defects but provokes an evolution of the nanostructure of the disorder. As seen in fig.\ref{fig.4}b, $\Delta\rho/\rho_{0} $ progressively deviates from curve-B  to eventually reach curve-C when $\rho_{0}^{2D}%
\sim8k\Omega/\square$. A clear insulating behavior is only observed when
$\rho_{0}^{2D}$ exceeds $\sim10k\Omega/\square\ $ (curve-D in fig.\ref{fig.3}-a).
The data displayed in fig.\ref{fig.3}-b for the two LSCO samples exactly coincide with
curve C. This is also found when analyzing
the data reported for LS$_{0.1}$CO$\ $with 2\%Zn \cite{Komiya} and for
LS$_{0.15}$CO$\ $ with 4\%Zn \cite{Fukuzumi}. We might then conclude that
although different types of disorder produce qualitatively similar behavior,
nanostructured disorder induces larger upturns in $\Delta\rho(T)/\rho_{0}$.

\section{Realistic phase diagram including disorder}
\begin{figure}
\centering
\includegraphics[width=8.5cm]{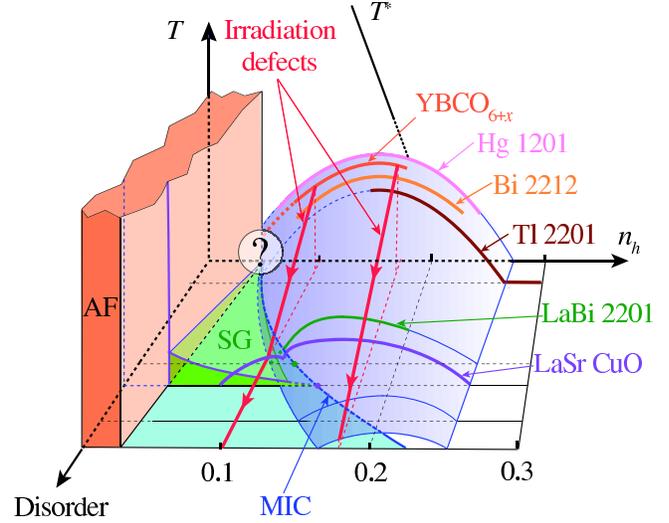}
\caption{Schematic phase diagram for high-$T_{c}$ cuprates as a
function of temperature $T$, hole doping $n_{h}$ and
disorder. The shaded orange volume represents the AF range
and the violet volume is the SC dome. The pseudogap line
is indicated and is not affected by disorder \cite{Alloul}. The small static
magnetically disordered spin glass like regime (SG) is sketched in green. 
The red lines show the trajectories followed by our underdoped YBCO$_{6.6}$ 
and optimally doped YBCO$_{7}$ samples under electron irradiation. The various systems are ranged with respect to their optimal $T_{c}$. The estimates
of $n_{h}$ are matched with an empirical correspondance with the
case of La$_{2-x}$Sr$_{x}$CuO$_{4}$ for which $n_{h}=x$ \cite{Presland},
except for La-Bi2201 where $n_{h}$ has been deduced from
thermopower measurements \cite{Ando2}. The dotted blue line indicates the doping at which a real metal-insulator crossover (MIC) can be ascertained (see text).}
\label{fig.5}
\end{figure}

The results obtained here allow us to propose in fig.\ref{fig.5} an extended 3D version
of the ($T,n_{h}$) phase diagram of the cuprates in which disorder is
introduced as the third axis parameter. While disorder
only affects moderately the AF phase, it depresses $T_{c}$ much more in
the underdoped than in the overdoped case \cite{Alloul, Kluge, Tallon},
hence the asymetric shrinking of the superconducting dome.  The $T_{c}$
trajectories followed in this 3D diagram for our electron irradiated
samples are shown by full red lines. As we have seen hereabove, the existence  of $\rho (T)$ upturns does not imply a $T=0$ insulating state, but is associated at low defect content with
inelastic scattering processes. The actual MIC (blue line in fig.\ref{fig.5}) takes place
when a divergence of $\rho (T)$ is detected, which only occurs for
sufficient disorder .

On this 3D representation we have ranged the ($T,n_{h}$) phase diagrams of
the various pure cuprate families with respect to their optimal $T_{c}$
values, which can be considered as a reasonable physical quantity
representative of the intrinsic disorder in these systems. The MIC locations for
La-Bi2201 and LSCO, determined as for the irradiated YBCO, are represented
here as full dots. We can see that the La-Bi2201 SC dome and MIC match
those induced by electron irradiation in YBCO. On the contrary, the fact that for LSCO the SC
dome only qualitatively fits in this diagram can be assigned to the different nanoscale disorder associated with the local stripe ordering favored for $x\approx0.12$. We have also sketched the SG phase which extends with disorder in a similar manner as the
MIC.
 
As for the  four families of rather clean cuprates Hg-1201,
YBCO, Bi2212 and Tl-2201, they have been ranged as well here with respect to
their optimal $T_{c}$, respectively 97, 93, 90 and $\sim 85$K \cite{Tyler}.
At this fine level, the ordering might not be representative of the residual
disorder as it does not take into account the influence of the number of CuO$%
_{2}$ layers (and their buckling). 

Let us further point out that  the representation done here is yet
somewhat simplified as a constant disorder is assumed throughout the phase
diagram for each cuprate family. This is less critical for the lower $T_{c}$ cuprate families than
 for the cleanest ones such as YBCO, where the hole content and chain disorder are fully
interconnected. The data of fig.\ref{fig.3}-c allow us to indicate that a MIC occurs for 
$x\simeq0.35$ for disordered chain
samples. For this oxygen content an AF phase is usually
detected. So, an essential consequence of our study is that the
intrinsic properties in this family are only accessible on reasonably well ordered
compounds such as YBCO$_{7}$, YBCO$_{6.6}$ or for the chain
ordered orthoII YBCO$_{6.5}$ used recently to probe the fermi surface by bulk
transport measurements \cite{Doiron}. 

\section{Conclusion}
Here we have used the progressive electron irradiation of the clean YBCO system to provide a true control of the disorder parameter for specific dopings in the 3D phase diagram that we propose. 
This has allowed us to demonstrate that the apparent MIC detected so far in cuprates is always driven by the existing disorder. We have thus unravelled the confusion brought about by the report of similar -$\log T$ dependences in the ''insulating'' state of all these compounds. Those are 
due for low defect contents to scattering processes intimately linked with the strong correlation effects and incipient magnetism. These results, together with the Curie Weiss like susceptibilities found by NMR for in plane defects display a strong similarity with Kondo effect in dilute alloys. Although the microscopic picture does differ, and no detailed theoretical justification has been proposed so far, the data can be analyzed in the frame of a Kondo like phenomenology. We have also shown that the magnitude of the -$\log T$ variation rather depends on the morphology of the disorder.

So the physical properties of most ''pure'' cuprates families are affected by their specific disorder, its influence being quite critical in the underdoped pseudogap phase, for which those electron correlation effects are most important. We have evidenced here that, even if the true microscopic character of the disorder is not fully characterized in a given cuprate family, the actual value of its optimal $T_{c}$ and the hole content at which a MIC is detected are direct measures of the disorder, that allow us to position the families on the 3D phase diagram.

Contrary to the proposal that
disorder and spin glass behavior is generic to the physics of the CuO$_{2}$
planes \cite{Panagopoulos}, an essential conclusion of our study is that the
intrinsic properties are only accessible on reasonably well ordered
compounds. The very shape of the ($T,n_{h}$) phase diagram for vanishing
disorder could correspond to a merging of the SC and AF lines as found in
other correlated electron systems such as heavy fermions and organic
superconductors. Such a speculation is illustrated in fig.\ref{fig.5}, by
the hypothetical trend of the 3D phase diagram for vanishing disorder.

\acknowledgments
We thank P. Lejay, D. Colson and A. Forget for providing the single crystals. This work was supported by the ANR grant NT05-4\_41913. The work at NHMFL was supported
by NSF and DOE. The experiments at the LNCMP have been financed by the
contract FP6 "Structuring the European Research Area,
Research Infrastructure Action" R113-CT2004-506239.


\begin{thebibliography}{0}
\bibitem{rem}
One cannot exclude that such a deviation might also exist for O$_{6.6}$-B 
as data acquisition was done above
2K in this experiment.
\end{thebibliography}
\end{document}